\documentclass[conference]{IEEEtran}
\IEEEoverridecommandlockouts
\usepackage{cite}
\usepackage{amsmath,amssymb,amsfonts}
\usepackage{algorithmic}
\usepackage{graphicx}
\usepackage{textcomp}
\usepackage{import}
\usepackage{xcolor}
\usepackage{cleveref}
\def\BibTeX{{\rm B\kern-.05em{\sc i\kern-.025em b}\kern-.08em
    T\kern-.1667em\lower.7ex\hbox{E}\kern-.125emX}}

\begin{document}

\title{Analytical Engines With Context-Rich Processing:\\
Towards Efficient Next-Generation Analytics
%Efficiently Shedding Light On Dark Data
}

\author{\IEEEauthorblockN{Viktor Sanca}
\IEEEauthorblockA{%\textit{DIAS Lab} \\
\textit{EPFL}\\
Lausanne, Switzerland \\
viktor.sanca@epfl.ch}
\and
\IEEEauthorblockN{Anastasia Ailamaki}
\IEEEauthorblockA{%\textit{DIAS Lab} \\
\textit{EPFL}\\
Lausanne, Switzerland \\
anastasia.ailamaki@epfl.ch}
}

\maketitle

\begin{abstract}
% general way something is done/problem/evolution
As modern data pipelines continue to collect, produce, and store a variety of data formats, 
%traditional analytical engines are usually limited to extracting insights from (semi-)structured data,
extracting and combining value from traditional and context-rich sources such as strings, text, video, audio, and logs becomes a manual process where such formats are unsuitable for RDBMS.
% how this problem is exacerbated/ why is it growing
To tap into the dark data, domain experts analyze and extract insights and integrate them into the data repositories. This process can involve out-of-DBMS, ad-hoc analysis, and processing resulting in ETL, engineering effort, and suboptimal performance. While AI systems based on ML models can automate the analysis process, they often further generate context-rich answers. Using multiple sources of truth, for either training the models or in the form of knowledge bases, further exacerbates the problem of consolidating the data of interest. 

% vision we propose
We envision an analytical engine co-optimized with components that enable context-rich analysis. Firstly, as the data from different sources or resulting from model answering cannot be cleaned ahead of time, we propose using online data integration via model-assisted similarity operations. Secondly, we aim for a holistic pipeline cost- and rule-based optimization across relational and model-based operators. Thirdly, with increasingly heterogeneous hardware and equally heterogeneous workloads ranging from traditional relational analytics to generative model inference, we envision a system that just-in-time adapts to the complex analytical query requirements.  
% why is it impactful
To solve increasingly complex analytical problems, ML offers attractive solutions that must be combined with traditional analytical processing and benefit from decades of database community research to achieve scalability and performance effortless for the end user.
\end{abstract}

\begin{IEEEkeywords}
analytics, machine learning, data integration, query optimization, hardware-conscious processing
\end{IEEEkeywords}

\section{Introduction}\label{sec:intro}
One of the main goals of data analytics is to extract insights, value, and intelligence from the available data sources. Traditionally, in the case of relational database management systems, the data would be stored in tables over which users declaratively specify \texttt{what} they would like to extract from the data, with several layers of abstractions performing various logical and physical optimizations. This takes the complexity, and the responsibility from the hands of the user and relieves them from specifying \texttt{how} the data processing pipeline should execute.

However, we are witnessing and participating first-hand in producing context-rich data, such as documents, text, images, and videos, through social media or other means. This trend is growing and expected to continue with the adoption of metaverse-like technologies~\cite{burgener_rydning_2022}. The underlying context, such as sentiment or interest detection from text or object recognition and classification from image sources, represents a rich source of value, however unreachable and therefore called \textit{dark} to the data processing tools. 

While traditional database systems lack the means to analyze such unstructured data, the explosion of research in machine learning has provided both algorithms and systems fueled by large amounts of data and the latest technology to enable super-human performance in various inference and classification tasks. Tasks previously requiring domain experts and significant intervention of the human-in-the-loop can take place automatically using artificial intelligence. Also, models such as GPT-3~\cite{DBLP:conf/nips/BrownMRSKDNSSAA20} can also represent data sources, generating new data. More broadly, Foundation Models~\cite{DBLP:journals/corr/abs-2108-07258} provide an opportunity to adapt large-scale models to specific tasks. Overall, machine learning provides us with analytical capabilities to process and transform the data with human-like context-rich semantics while performing at super-human performance, concerning the speed of processing and the quality of the result.

\begin{figure}
    \centering
    \includegraphics[width=\linewidth]{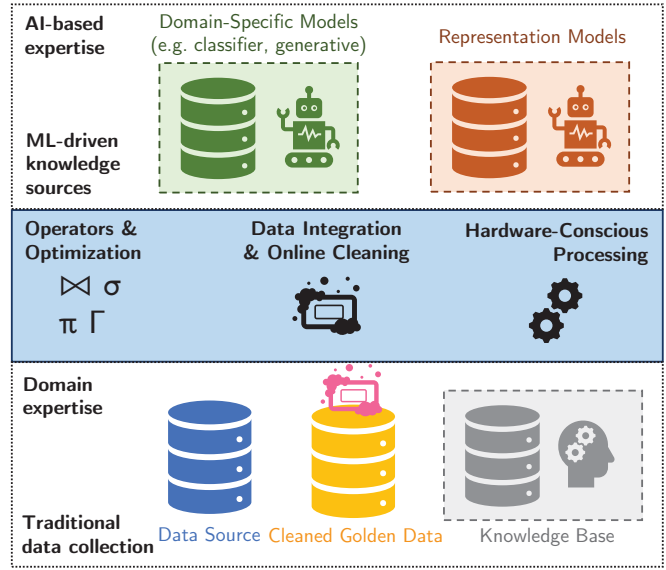}
    
    \caption{Analytics with traditional and model-driven sources and processing.}
    \label{fig:conceptual}
  
\end{figure}

We group data and insight sources into two categories in~\Cref{fig:conceptual}. The upper part depicts the sources whose analysis is ML-driven and often have their software systems run individually. Similarly, we consider a traditional data source structured data, collected from various sources (sensors, transactions, domain experts) and cleaned to a certain extent. In a complex data processing pipeline that needs to combine two sources, the user is faced with complex decisions, departing from declaratively specifying only \texttt{what} they would like to achieve. Such manual processing is tedious, error-prone, or inefficient and would require a high level of system- and programming- in addition to domain-expertise. To further exacerbate the issue, manually tuning the systems for the particular query to run on increasingly heterogeneous hardware is difficult and impractical. 

The challenge for performing optimizations and combining data from various sources (either collected or trained on) is that the data cannot be fully cleaned and unified to an expected schema and values ahead of time. Furthermore, the source of dirty data is less likely to be a mistake such as misspelling but a word with the same semantics (synonym, alternative spelling, alternative forms) - further representing a context-rich online data cleaning task.  

%What database systems could further do to reduce overlapping and repetitive computations, is to progressively cache prior query intermediate results, such as in the pipeline recycling paper \cite{DBLP:conf/sigmod/IvanovaKNG09}, after model invocation or after similarity matching. 

To outline the design goals and integration challenges, opportunities, and potential solutions to integrate context-rich processing with analytical engines (the blue-shaded part of \Cref{fig:conceptual}), in the rest of the paper, we describe:
\begin{itemize}
    \item a motivating example in~\cref{sec:motivation},
    \item methods for processing context-rich data in~\cref{sec:processing},
    \item integration of the context-rich data sources expert-free and on-the-fly in~\cref{sec:qp},
    \item exposing logical and physical optimizations in~\cref{sec:qo},
    \item the challenge of optimizing for hardware heterogeneity in~\cref{sec:hw}.
\end{itemize}

We envision an analytical engine that declaratively combines context-rich processing with traditional data sources to hide from the user the complexity of logical and physical optimization, underlying hardware, and resulting on-the-fly data integration. 

%1.5p

\section{Motivation}\label{sec:motivation}
\begin{figure}
    \centering
    \includegraphics[width=\linewidth]{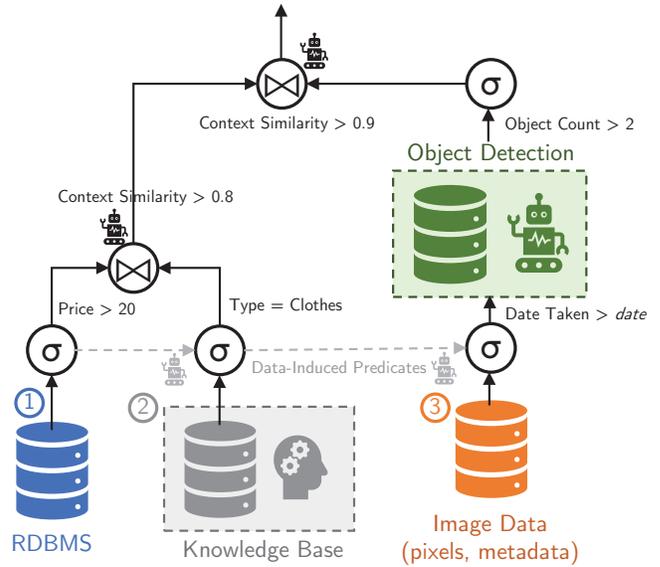}
    %\vspace{-18pt}
    \caption{A query that combines both context-rich and traditional analytics.}
    \label{fig:query}
    %\vspace{-12pt}
\end{figure}

Suppose an online shopping platform wants to analyze the data and trends using several sources:
\begin{enumerate}
    \item traditional RDBMS where product, user, and transaction details are managed and stored, 
    \item a general knowledge base to supplement and extend the product information based on domain expertise,
    \item image storage of the products (from reviews, other websites, or social media).
\end{enumerate}

The data analyst wants to answer which \textit{clothing} products with a price greater than 20 appear in customer images taken after a specific \textit{date}, but such that it is not only the picture of the clothing apparel but other objects appear too - possibly indicating that it is not a generic product description image. This query is conceptually represented in~\Cref{fig:query}.

There are three distinct tasks that a data analyst has to perform:
\begin{enumerate}
    \item extract the qualifying data from RDBMS, 
    \item infer which ones are clothing items and combine the results with matching clothing objects in images,
    \item analyze the photos taken after a specific date if they contain more than two objects.
\end{enumerate}

With an RDBMS, step 1) is entirely declarative, and the user can retrieve and export the qualifying records efficiently. However, if not careful, the user may not push down the filter in the knowledge base to reduce the qualifying records for the first join, where only clothing items should be selected. Furthermore, the knowledge base may have been curated and collected on a different and broader dataset that does not precisely match the labels or available data in the RDBMS. These differences are more likely to occur due to synonyms or some other alternative word or phrase forms that may not be resolvable with traditional string similarity algorithms and where the user may need to manually or with the help of an expert resolve which items are indeed clothing items, which can be both error-prone and heavily involve human in the loop which increases the effort and execution time. Finally, the analyst may make a mistake, and instead of first filtering out images after a specific date, perform heavy processing on all the corpora and similarly have issues merging the object labels with ones coming from steps 1) and 2). 

Overall, the analyst now has the task to carefully orchestrate possibly an ad-hoc query and multiple sources of data and system components. The cost of logical or physical optimization mistakes, as well as not considering scalability and potentially not processing data beyond main-memory capacity, can quickly hinder even a reasonably simple context-rich query and make it frustrating to use and design such pipelines, abandoning the potential value of the data. Furthermore, every query becomes a hand-tuned orchestration between the components, including logical optimization steps, physical optimization, and hardware resource management, despite all the lessons learned from decades of data management research.

Several requirements surface from the example, reiterating the need for declarative processing where users do not need to specify \texttt{how}, but only \texttt{what} they want to process. To have a declarative context-rich analysis system, we propose a design that necessitates the following:
\begin{enumerate}
    \item cost- and rule-based optimization, 
    \item hands-free hardware resource management,
    \item ad-hoc data matching and integration,
    \item exposing automated context-rich processing methods. 
\end{enumerate}

The broad database and machine learning research community have focused actively on solving some or all of these issues. In the next section we will discuss the need for automated context-rich data analysis methods.

%0.5p   %2p

\section{Analyzing Context-Rich Data}\label{sec:processing}
We broadly use the term context-rich data for all the data sources that would require a human or a domain expert to infer the meaning (context) that is not immediately stated as machine-understandable rules. For example, string edit distance or locality-sensitive hashing-based string similarity can compare strictly specified characteristics, but such methods cannot capture string synonyms in a straightforward way, where one would use a dictionary or a human-in-the-loop. Similar can apply to images, where finding the most prevalent colors is easy to specify and compute; however, understanding the image setting and performing object detection and classification would be considered context-rich analysis.

One way to perform the such large-scale analysis is to deploy crowdsourcing approaches such as Amazon Mechanical Turk\cite{amazon_mechanical_turk} or reCAPTCHA\cite{von2008recaptcha}, or employ domain experts. These approaches have proven invaluable in annotating data and create sources that were, for example, immensely useful in training machine learning models. However, the human-based analysis would be error-prone, slow, and expensive for analytics.

On the other hand, machine learning had its renaissance due to the data volume explosion, hardware processing power improvement, and novel systems and algorithms, which enabled creating of tools that outperform humans even at context-rich processing tasks. Tasks such as text, image, audio, and video analysis are ongoing topics of research, and specific tasks yield even super-human performance. More importantly, obtaining high-quality models trained on Web-scale data as a commodity resource that can be modified via mechanisms such as one- or few-shot learning is possible, for example, via the idea and initiative of Foundation Models~\cite{DBLP:journals/corr/abs-2108-07258}. Context-rich processing must happen seamlessly and scale with the data and request volume, and model-based approaches allow high-recall analysis on various tasks.

It is desirable to use models pre-trained on web-scale data with significant compute resources and fine-tune them to the particular task or dataset at hand. Creating and using models and running them efficiently is possible through frameworks such as Tensorflow~\cite{tensorflow2015-whitepaper} or PyTorch~\cite{NEURIPS2019_9015}, with further extensions such as Keras~\cite{chollet2015keras} to further make models more declarative.

There has been a significant interest in bringing together statistical processing with ML training and inference into the traditional data management systems, using systems such as Raven~\cite{DBLP:conf/cidr/KaranasosIPSPPX20}, MADLib~\cite{10.14778/2367502.2367510}, or SystemML~\cite{10.14778/3007263.3007279} to perform inference or learning on tasks such as regression, classification, clustering to extract richer insights from existing, often tabular data. The lessons and the need are the same: we need viable integration for context-rich analysis and logical and physical optimizations for new data sources and value to avoid the pitfalls of manually constructing complex unoptimized data processing pipelines.

Since the models have been trained on large-scale data, and conversely, we can use data to enrich the analytical pipeline from various sources such as knowledge bases, we cannot assume that data is integrated and cleaned even in the outputs. Cleaning the data ahead of time is impractical, if not impossible, due to the scale. Furthermore, models are probabilistic and do not produce only labels (which can be context-rich on their own), for example, as in \Cref{tab:sem_example}, but generative models can produce output and data on their own~\cite{DBLP:conf/nips/BrownMRSKDNSSAA20}. Integrating the data between the analytical pipelines becomes a key challenge since online result consolidation typically remains the only option~\cite{DBLP:conf/sigmod/GiannakopoulouK20}.

\begin{table}[htbp]
%\vspace{-10pt}
\caption{Example of context-rich text labels that models may output}
\begin{center}
\begin{tabular}{|c|c|}
    \hline
    category & semantic matches \\ 
    \hline
    dog & dog, canine, golden retriever, puppy \\
    cat & cat, maine coon, feline, kitten \\
    animal & cat, dog, golden retriever, feline \\
    \hline
    shoes & boots, sneakers, oxfords, lace-ups \\
    jacket & blazer, coat, parka, windbreaker \\
    clothes & boots, parka, windbreaker, coat \\
    \hline
\end{tabular}
\label{tab:sem_example}
\end{center}
%\vspace{-6pt}
\end{table}

However, learned approaches allow us to observe this as a representational learning problem that would allow automated resolution, grouping, joining, and filtering based on context-similarity (such as synonyms and similar categories). Approaches such as word2vec~ and\cite{DBLP:journals/corr/abs-1301-3781}, fastText~\cite{DBLP:journals/tacl/BojanowskiGJM17} create a latent representation of strings to learn word associations from large corpora of text, such as Wikipedia~\cite{wiki} or Common Crawl~\cite{common_crawl}. Such approaches can match similarities based on synonyms, semantics, alternative spellings, different tenses, and even misspellings~\cite{DBLP:conf/naacl/EdizelPBFGS19}. Context-based filtering, joining, and grouping based on similarity is then converted to finding similarity between original data representation in the latent, high-dimensional vector space - a task that can be specified and automated.

In the next section, we will discuss how using models to analyze the data and combining the output from different parts of the query can be expressed as operators amenable to logical and physical optimizations across different data processing pipelines.
         %1p     %3p

\section{Query Processing And Data Integration}\label{sec:qp}
\begin{figure}
    \centering
    \includegraphics[width=\linewidth]{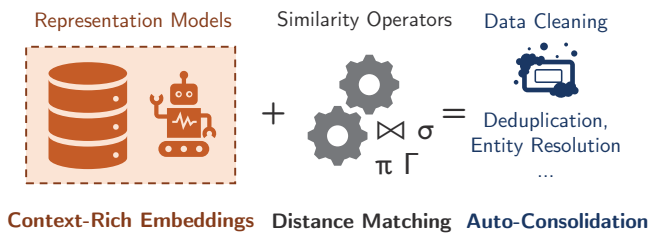}
    %\vspace{-18pt}
    \caption{Automated, on-the-fly result consolidation.}
    \label{fig:clean}
    %\vspace{-12pt}
\end{figure}

To enable a variety of logical and physical optimizations of the holistic query pipeline, no matter where standard relational or model-based operators are placed, we need to start from a common intermediate representation (IR). Approaches such as Weld~\cite{DBLP:journals/pvldb/PalkarTNTPNSSPA18} or Raven~\cite{DBLP:conf/cidr/KaranasosIPSPPX20} already take significant steps towards a common representation that is amenable to optimization rules.

Since we are to use external models in the pipeline, we need to express some properties of context-rich analysis operators. 
We need to specify the expected input and output schema at the logical level. Furthermore, we need to select the transformations such operator applies on the input, in terms equivalent to the selection, projection, union, or joins, and what effect operator push-down/pull-up has on other, for example, relational operators. Finally, in this step, we must include high-level cost information, such as the effect on the input/output cardinality. This ensures semantic equivalence and correctness and can be expressed in existing query optimization frameworks~\cite{DBLP:conf/sigmod/BegoliCHML18}. 

Next, we consider some physical optimizations related to context-rich operators using models.
While some models perform computationally-intensive inference only, there are still models that perform similarity search or other methods which impose different costs based on data access patterns that we need to enumerate.
For example, index-based access for similarity search~\cite{johnson2019billion} should be accounted for in the optimization process in a cost-based optimization process. 

Finally, we must process and optimize by considering the potential polystore situation where we will have multiple sources of data and multiple processing engines. To add to this complexity, we also have increasingly heterogeneous hardware at our disposal (CPU, GPU, TPU, FPGA, ASICs, DPU, etc.), all with corresponding data transfer, processing, and startup costs. We need to annotate the processing unit or operator traits and perform decisions that adapt to the underlying hardware, with principles described in the thesis of Chrysogelos on efficient analytical processing for CPU-GPU platforms~\cite{Chrysogelos:296204}. Just-in-time decisions are needed to make correct decisions in growing hardware, operator, and system heterogeneity. To fully use the underlying hardware capabilities, this merits optimizing novel analytical operators individually for existing or new platforms and holistically for hybrid execution over a variety of compute units.

% result integration 
Irreplaceable components of context-rich analysis are context-rich \textit{semantic operators} that can process, transform, and integrate data with traditional analytical operators. This task resembles data cleaning but has to capture context and semantics, perform the task efficiently at query time, and run without user supervision in an automated way. In particular, consider the example in \Cref{tab:sem_example}: if we want to join clothing items, without strong domain expertise or knowledge-base, we would not be able to perform matching, and still, data sources may contain synonyms, spelling differences, or other expert-system-detectable differences. To resolve this issue, we propose model-assisted context-rich processing that we call \textit{semantic join}, which we denote  with a small robot in \Cref{fig:query}. 

In practice, we can use representational models such as fastText~\cite{DBLP:journals/tacl/BojanowskiGJM17} to capture the context even in place of spelling mistakes~\cite{DBLP:conf/naacl/EdizelPBFGS19}. More complex models have the capacity to specialize in performing this task with more complex input, such as sentences~\cite{DBLP:conf/naacl/DevlinCLT19, DBLP:conf/nips/BrownMRSKDNSSAA20}, and work with more complex data, such as images or audio. In particular, model-assisted operators transform the input to latent, high-dimensional vector space, where distance-based metrics such as cosine similarity capture features such as context similarity~\cite{DBLP:journals/corr/abs-1301-3781}. We represent this in \Cref{fig:clean}, where a tedious and domain-expert task becomes completely automated, allowing on-the-fly result consolidation based on context.

Traditional techniques such as locality-sensitive hashing or other distance metrics capture syntactic similarity (such as misspellings); however, previously mentioned deep-learning methods learn the context from large data corpora. However, other optimizations known to speed up similarity operations, such as joins, are still applicable and should be explored in this context, where initial embeddings are already provided instead of created based on transformations such as hashing.

% new operators
We propose the following operator extensions:
\begin{itemize}
    \item \textbf{Semantic Select} - context-based filtering, e.g.\\ \texttt{word = "Clothes" using model "M" with cosine threshold >= 0.9},
    \item \textbf{Semantic Join} - joining two or more relations based on join key context, for example based on model latent space distance between the join key mappings,
    \item \textbf{Semantic GroupBy} - on-the-fly clustering of the result based on a model-based similarity threshold.
\end{itemize}

With semantic operators, more complex optimization techniques that work for relational data, such as data-induced predicates, can be evaluated and applied in the query plans~\cite{DBLP:journals/pvldb/OrrKC19}. The key idea is to make the specified unified query plan amenable to traditional relational plan optimizations.

% language and specification
We note that SQL may not be the best or the only way to represent such query plans; however having an optimizable representation is required to systematically perform a variety of transformations that preserve the correctness of the execution. Systems such as Weld\cite{DBLP:journals/pvldb/PalkarTNTPNSSPA18}, or effort to push Pandas down the DBMS~\cite{DBLP:conf/cidr/HagedornKS21} as a common data science tool - should be considered in such a next-generation system to keep it familiar and interoperable with existing libraries and code-ecosystem. 
      %1p   %4p

\section{Query Optimization}\label{sec:qo}
\begin{figure}
    \centering
    \includegraphics[width=\linewidth]{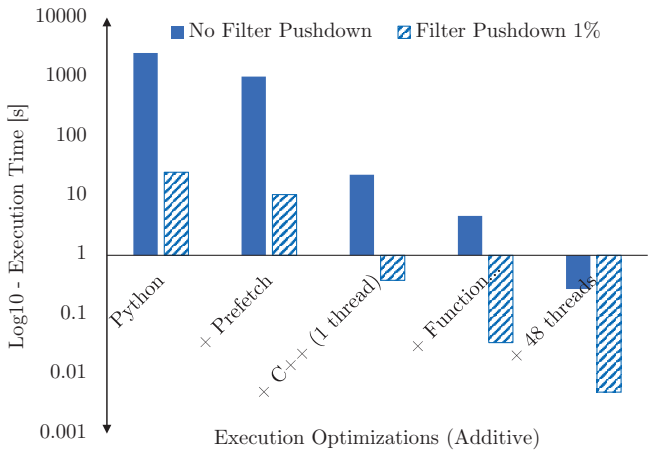}
    %\vspace{-18pt}
    \caption{Additive effects of logical and physical optimizations (log-scale).}
    \label{fig:optimizations}
    %\vspace{-12pt}
\end{figure}

Cost- and rule-based optimizations must be considered to produce a scalable, efficient, and hardware-conscious data-analytics pipeline. Furthermore, selecting the correct physical operator implementation and scaling the solution to available compute resources improves hardware utilization and end-to-end execution time.

We briefly analyze a model-assisted semantic similarity join to demonstrate the importance of optimizations. To simulate the operation, we join two arrays of 10k strings taken randomly from the Wikipedia dataset. We use fastText word embeddings with a dimension of 100, perform the similarity join with cosine distance where the threshold has to be greater than 0.9, and run the experiment on a server with two socket Intel Xeon CPU Gold 5118 CPU (2 x 12-core, 48 threads) and 384GB of RAM. We summarize the execution time and the effects of optimizations in \Cref{fig:optimizations}.

Suppose the data analyst takes the first tool at their disposal and uses Python: they load the data eagerly (that may not fit in memory ordinarily) and then specify the task using the available libraries to iterate over two loops and perform a similarity check. Suppose this is a part of a more complex pipeline, and the user did not apply the filter before processing - a standard filter pushdown rule. This represents the left-most bar, and the variant with 1\% selectivity applied beforehand in the blue patterned bar. This optimization already improves orders of magnitude even in Python in a single-threaded execution context. Since fastText produces a hash table of known words, we can further try to optimize the amount of data access by prefetching necessary data, a physical optimization detail that the user may not be aware of. The next step takes the execution to tighter code, and fewer function calls between libraries using C++, tightening the similarity code using CPU-specific instructions, and scaling up the execution. The effect of well-known optimizations in the database community brings many orders of magnitude improvements from the initial manual, intuitive solution. Conversely, applying such optimizations is tedious and error-prone if done manually; however, the frustration when waiting for an answer that takes thousands of seconds which can take less than a second may cause the user to abandon such analysis.

Inefficiencies can quickly add up in a more complex plan, as does the complexity of manually optimizing and tuning ad-hoc context-rich queries. Intuitively, performing expensive model inference or processing benefits equally, if not more, from correct join orders and filter pushdowns that potentially reduce the input cardinality. 

% how do we reason about context-rich operator cardinality
Exposing all the operators to optimizations and how they affect cardinality and the input/output characteristics is a necessary prerequisite for the optimization. Frameworks such as Apache Calcite~\cite{DBLP:conf/sigmod/BegoliCHML18} already support flexible operator characteristics, traits, and transformations. We could encapsulate such operators in a UDF-like manner while exposing details such as compute requirements, amenability to parallelizing the input, and memory and data transfer requirements to the optimizer component. Often, there are already highly optimized runtimes to perform inference~\cite{tensorflow2015-whitepaper} that can reduce development time and be fused more tightly in global processing orchestration. However, we have to be mindful that some models that are useful for context-rich analysis, such as word embedding models, have both inference modules and data structures such as hash tables and index structures~\cite{johnson2019billion} for expediting operations such as similarity or top-k searches, which have to be included in the optimization process equally as relational data indexes are.   

The key to optimizing a complex context-rich query, such as one in \Cref{fig:query}, is to provide correct abstractions that perform the optimizations while enabling the user to declaratively and in an intuitive way specify the data processing task at hand, possibly even making well-educated or learned decisions on some of the semantic operators without explicit user control.

%Speculation \cite{DBLP:conf/cidr/SioulasSMA21}
%Sampling \cite{DBLP:conf/damon/SancaA22}
   %0.5p   %4.5p

\section{Hardware-Conscious Optimizations}\label{sec:hw}  
Machine learning workloads were one of the primary drivers of mainstream GPU adoption, which was consequently explored in the context of data management on heterogeneous hardware. Increasingly heterogeneous hardware, with various computational, data movement, and storage access overheads, is a source of active research. The general problem becomes how to efficiently scale up the individual algorithms, tune them to hardware capabilities, orchestrate hybrid execution models, and schedule many machine-learning inference runtimes on the available hardware. 

\begin{figure}
    \centering
    \includegraphics[width=0.6\linewidth]{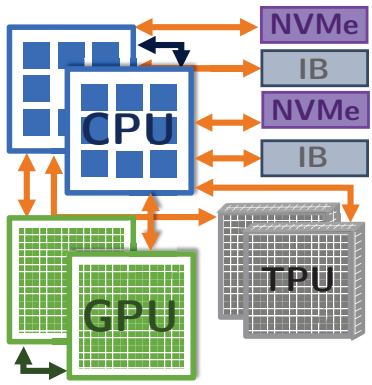}
    \caption{The increasing hardware heterogeneity is challenging to optimize for.}
    %\vspace{-12pt}
    \label{fig:hw_het}
\end{figure}

We represent the conceptual heterogeneous hardware layout in~\Cref{fig:hw_het}: it includes multi-core CPUs on multiple sockets, potentially each with a separately interconnected GPU. Furthermore, a TPU-like device is designed to perform fast and efficient inference~\cite{DBLP:conf/isca/JouppiYPPABBBBB17}, however, that can equally be considered as an accelerator for relational query processing~\cite{DBLP:journals/pvldb/HeNBSSPCCKI22}. It is worth noting that specialized hardware units for processing real-time neural network inference are becoming commonplace even in mobile phones, such as Apple Neural Engine or similar~\cite{neural_proc}. Finally, there are fast network (e.g., InfiniBand) and storage (NVMe) solutions, all interconnected with PCIe or other technologies. The big challenge is how to provision these resources correctly, how to place, split, and schedule the execution such that the increasingly complex processing pipeline efficiently uses the available resources, and for example, which lessons from hybrid CPU-GPU processing \cite{Chrysogelos:296204} we can take and further generalize.

With increasingly difficult cost and cardinality estimation, fast sampling running on modern hardware~\cite{DBLP:conf/damon/SancaA22} or speculation techniques~\cite{DBLP:conf/cidr/SioulasSMA21} can come in handy to provide mechanisms for practical and adaptive query optimization and execution.

Late binding to the query requirements concerning hardware and data layout requirements has become a standard in modern analytics. Just-in-time code generation using frameworks such as LLVM enables specializing the code paths and equally adapting to an increasing variety of data input and output formats. This is increasingly important, as we now consider not only typical relational formats but repositories that contain context-rich data such as strings, text, audio, images, and video, where the NoDB paradigm~\cite{DBLP:conf/sigmod/AlagiannisBBIA12} and adapting to data format heterogeneity at runtime~\cite{DBLP:journals/pvldb/KarpathiotakisA16} play a crucial role in handling lazy data access to a variety of data formats. Just-in-time code generation allows this to be specified as late as query runtime: for example, if model operator output specification is not known ahead of time, this can be detected, and the code path generated after the model outputs first data - with corresponding overheads and delay.

% we need to take into account the data sizes, and caching opportunities, late materialization etc for transfers, and the characteristics of the interconnect as well
% also, specific instructions and compute that may be available

Data layouts, data size, compression effects, and access method selection become a factor to consider. For example, recent research has yielded specialized databases for vector processing~\cite{DBLP:conf/sigmod/WangYGJXLWGLXYY21}. Integrating complex pipelines through a well-coordinated polystore mechanism or a fully generated pipeline should take different data formats as a first-class problem. For example, the effects of cache hierarchy or hardware-specific vectorized instructions can significantly change the behavior of processing a high-dimensional vector. Furthermore, optimization opportunities such as inference using hardware-enabled half-precision (or lower) floating point formats need to be considered and specialized for~\cite{DBLP:journals/corr/abs-1805-06085}. 

Finally, data movement and compute shipping have to be carefully designed. Obvious questions such as data compression before sending the data over the interconnect for processing come to mind. However, we should also consider the cost of shipping and initializing model parameters and runtime on devices as a potential cost and optimization, as complex models can have many millions of parameters~\cite{DBLP:journals/corr/abs-2108-07258}. The decision of the concrete operator execution happens in a managed runtime such as Tensorflow~\cite{tensorflow2015-whitepaper} or being optimized entirely using a relational pipeline~\cite{DBLP:conf/cidr/KaranasosIPSPPX20}, further opening up the landscape of scheduling and optimizing complex data-science pipelines to use available heterogeneous hardware resources adaptively based on query-specific requirements.

% \noindent\fbox{%
%     \parbox{\linewidth}{%
%         \textbf{Challenge: specify the challenge}\\
%         Some text
        
%         \textbf{Vision:}\\
%         text
%     }%
% }
          %1p     %5.5p

\section{Related Work}\label{sec:related}
The tasks analytical engines are supposed to support and optimize are evolving with novel algorithms, sources of values, applications, and hardware platforms. The research community and industry are proposing systems and investing significant research efforts to address them. We can use them as foundation blocks for next-generation analytical engines for context-rich processing.

\subsection{Querying and Optimizing Complex Analytical Pipelines}
To extract deeper insights from the data, more complex data transformations using User Defined Functions (UDFs) or external libraries can take place. In our case, context-rich analysis can happen as a UDF or by invoking another framework optimized to perform specific tasks (e.g., image object detection). A mounting challenge is optimizing the external operators in the query, where systems like Froid~\cite{DBLP:journals/pvldb/RamachandraPEHG17} optimize UDFs for Relational DBMS, Weld~\cite{DBLP:journals/pvldb/PalkarTNTPNSSPA18} reduces the cost of data movement through optimizing a standard IR, Raven~\cite{DBLP:conf/cidr/KaranasosIPSPPX20} optimizes ML inference with an RDBMS, while Caesar~\cite{DBLP:conf/cidr/HerlihyCA22} aims to minimize the overheads between application boundaries. MADLib~\cite{10.14778/2367502.2367510} or SystemML~\cite{10.14778/3007263.3007279} allow training and inference over the input, often tabular data. Frameworks such as TensorFlow perform dataflow optimizations for model training and inference~\cite{10.1145/3088525.3088527}. We envision a system not limited to using models representing only the data in RDBMS but can cross-optimize and use the models from different data sources in a polystore-like design. Our vision aims to combine insights and value from, for example, relational data, knowledge bases, and images in a single analytical pipeline and holistically optimize the analytical pipeline across models and frameworks.

Rather than specifying the queries using languages such as SQL, using a standard Python data analysis library like Pandas and the work on how to expose it to RDBMS optimizations~\cite{DBLP:conf/cidr/HagedornKS21} could be a good step to improve usability and reduce the potential complexity of specifying new operators that would be needed for context-rich analytical engine. The end goal would be taking the existing code in languages such as Python and optimizing it automatically by creating a standard IR, detecting physical and logical optimizations, and correctly tuning the operators to the underlying hardware capabilities, computation optimizations, and data transfers.

\subsection{System Heterogeneity and Data Integration}

The analytical engine components that can perform context-rich analysis are part of reality and exist in individual components. Our vision is to bring them together in an efficient engine that would relieve users from performing tasks other than declaratively specifying what they would like to analyze in the data.

Foundation Models~\cite{DBLP:journals/corr/abs-2108-07258} offer a way to democratize models that would be practically impossible to retrain individually due to the required data and computation scale. However, such models can be specialized for various tasks to perform better than a human counterpart on context-rich data. Representational Models such as word2Vec\cite{DBLP:journals/corr/abs-1301-3781} and fastText\cite{DBLP:journals/tacl/BojanowskiGJM17} allow learning the word context and reasoning about such abstract similarity by embedding the text and performing similarity operations in latent space.

Since some of the models are already trained on web-scale data, and the analyst will likely supplement the knowledge with external data sources such as knowledge bases, performing data cleaning tasks on all the input data is impractical, if not impossible. However, data cleaning remains crucial in integrating data from different sources. To this extent, we need to incorporate relevant data in an online rather than ahead-of-query offline process through a query-driven mechanism previously explored in the context of repairing functional dependency violations~\cite{DBLP:conf/sigmod/GiannakopoulouK20}.

\subsection{Adapting to Data and Hardware Heterogeneity}
New applications, including machine learning, drive an increasingly heterogeneous hardware landscape. From the CPU-dominant landscape, we have a variety of compute units such as CPU, GPU, TPU, DPU~\cite{DBLP:conf/hotchips/Burstein21}, FPGA, ASICs, and others - not only on high-end servers but also on commodity devices such as mobile phones, such as Apple Neural Engine~\cite{neural_proc}. This is supplemented with various storage devices and interconnects that add to the opportunities for hybrid execution on multiple hardware components and novel optimizations for out-of-core algorithms. 

To adapt to the hardware heterogeneity and complex requirements, analytical engines such as Hyper~\cite{DBLP:journals/pvldb/Neumann11}, NoisePage~\cite{DBLP:conf/cidr/PavloAALLMMMPQS17}, and Proteus~\cite{DBLP:journals/pvldb/ChrysogelosKAA19} deploy just-in-time decisions via code generation using frameworks such as LLVM~\cite{lattner2004llvm}. On the side of machine learning, systems such as TensorFlow use JIT-compilation and specialize the runtime graph utilizing an approach called XLA using a domain-specific compiler.

There has been a significant effort in developing optimization frameworks, such as Apache Calcite~\cite{DBLP:conf/sigmod/BegoliCHML18} for performing logical, rule- and cost-based optimization over heterogeneous data sources. Furthermore, at a lower level of optimization, frameworks such as MLIR~\cite{mlir} provide a hybrid IR which can aid in bringing together different system and software components and optimizing them for heterogeneous hardware under unified infrastructure.

         %0.75p     %6.25p

\section{Towards Next-Generation Analytics}\label{sec:conclusion}    
%problem
In addition to the typical structured data analytics, humans produce and collect data that can contain value locked behind rich context for a traditional analytical engine. This may further get exacerbated by proposed use cases such as metaverse, where combining analytical insights from sources of interaction such as images, text, audio, and traditional data sources presents an untapped source of insight.

%proposal
While context-rich analysis, such as object detection for images or sentiment analysis in text, has proposed solutions using machine learning techniques, these solutions are often optimized individually. For more complex analytical pipelines, the user would perform physical, logical, and hardware optimizations for potentially ad-hoc analytical queries, which we want to avoid through a unified, next-generation analytical engine that applies the lessons learned in data management.

%impact
With many specialized and siloed solutions, both ML and DB communities make strides to enable richer and more efficient solutions to extract insights and value from data. We propose a system that performs holistic data processing pipeline optimizations merges the potentially unexpected or dirty results on the fly, and relieves the user from making decisions over the increasingly complex system and hardware landscape.

%outlook
%We envision a next-generation analytical engine that generalizes analytical operations over novel data analysis methods while achieving the performance of individually executing tasks in specialized and tuned components through just-in-time adaptation.
% MORE CONCRETE - BE CONCLUSIVE
We envision a next-generation analytical engine that generalizes analytical operations driven by new data processing methods to enable novel ways to extract and combine insights from different sources. At the same time, users are given control of data integration through a new class of semantic and model-assisted relational operators that enable automated result consolidation. Through vertical optimizations, correct abstractions, and just-in-time adaptation, a single declarative framework aims to reach the performance of individually and manually tuned components.
      %0.25p   %6.5p

%\section*{Acknowledgment}
%\nocite{*}

\bibliographystyle{IEEEtran}
\bibliography{references.bib}

\end{document}